\documentclass[useAMS,usenatbib]{mn2e}
\usepackage{amssymb}
\usepackage{graphicx}
\usepackage{xspace}

\usepackage{amssymb,latexsym,graphicx,natbib,eufrak,times,amsmath}

\usepackage{ifthen}
\def\draftversion{0} 

\makeatletter
\newcommand\mytoc{%
    \@starttoc{toc}%
}
\makeatother

\ifthenelse{\equal{\draftversion}{0}}{
	\usepackage{xcolor}
	\newcommand{\tmp}{}
	\newenvironment{envcomm}[1]{\renewcommand{\tmp}{#1}\begin{color}{blue}\begin{center}\hrule\vspace{0.5mm}\tmp's COMMENTS\end{center}}{\begin{center}END OF \tmp's COMMENTS\vspace{0.5mm}\hrule\end{center}\end{color}}
	\newenvironment{draft}{\begin{color}[rgb]{0,0.4,0}\begin{center}\hrule\vspace{0.5mm}DRAFT\end{center}}{\begin{center}END OF DRAFT\vspace{0.5mm}\hrule\end{center}\end{color}}
	\newcommand{\comcomm}[2]{\begin{color}{blue}\ $\bullet$ \textbf{#1:} #2 $\bullet$\ \end{color}}
	\newcommand{\revend}[1]{\par\begin{color}[rgb]{0,0.4,0}\begin{center}\hrule\vspace{0.5mm}END OF #1's REVISIONS\vspace{0.5mm}\hrule\end{center}\end{color}\par}
	\newcommand{\todo}[1]{\begin{color}{red}\ $\bullet$ \textbf{To do: }#1 $\bullet$\ \end{color}}
	
	\newcommand{\del}[1]{\begin{color}[rgb]{0,0.5,0.0}\ $\bullet$ \textbf{Deleted: }#1 $\bullet$\ \end{color}}
	\newcommand{\sk}[1]{\begin{color}[rgb]{0.6,0,0.6}#1\end{color}}
	\newcommand{\toc}{\par\begin{color}[rgb]{0.6,0,0.6}\begin{center}\hrule\vspace{0.5mm}\begingroup\small\let\cleardoublepage\relax\let\clearpage\relax\mytoc\endgroup\vspace{0.5mm}\hrule\end{center}\end{color}\par}
	}{
	\newsavebox{\trashcan}
	\newenvironment{envcomm}[1]{\begin{lrbox}{\trashcan}\begin{minipage}{\columnwidth}}{\end{minipage}\end{lrbox}}
	
	\newcommand{\comcomm}[2]{}
	\newcommand{\revend}[1]{}
	\newcommand{\todo}[1]{}
	
	\newcommand{\del}[1]{}
	\newcommand{\sk}[1]{}
	\newcommand{\toc}{}
	}

\long\def\symbolfootnote[#1]#2{\begingroup%
\def\thefootnote{\fnsymbol{footnote}}\footnote[#1]{#2}\endgroup} 

\newcommand{\aj}{AJ}
\newcommand{\araa}{ARA\&A}
\newcommand{\apj}{ApJ}
\newcommand{\aap}{A\&A}
\newcommand{\mnras}{MNRAS}
\newcommand{\pasp}{PASP}
\newcommand{\pasj}{PASJ}
\newcommand{\nat}{Nature}

\newcommand{\bb}[1]{\ifmmode \mbox{\boldmath $ #1$} \else  \mbox{\boldmath $#1$} \fi}

\newcommand{\U}[1]{\ensuremath{\mathrm{~#1}}}

\newcommand{\Myr}{\U{Myr}}
\newcommand{\Gyr}{\U{Gyr}}
\newcommand{\pc}{\U{pc}}
\newcommand{\kpc}{\U{kpc}}
\newcommand{\Mpc}{\U{Mpc}}
\newcommand{\msun}{\U{M}_{\odot}}
\newcommand{\Msun}{\msun}

\newcommand{\kms}{\U{km\ s^{-1}}}

\newcommand{\nbtt}{\texttt{NBODY6tt}\xspace}
\newcommand{\emacss}{\texttt{EMACSS}\xspace}

\newcommand{\nbody}{\texttt{NBODY6}\xspace}

\newcommand{\fig}[2][]{Figure#1~\ref{fig:#2}}

\newcommand{\sect}[2][]{Section#1~\ref{sec:#2}}

\renewcommand{\fig}[2][]{Fig#1.~\ref{fig:#2}}

\title[Secular galactic growth on star clusters]{The effect of secular galactic growth on the evolution of star clusters}

\author[Renaud \& Gieles] {Florent~Renaud\thanks{f.renaud@surrey.ac.uk} and Mark~Gieles\\
Department of Physics, University of Surrey, Guildford, GU2 7XH, UK
}

\date{Accepted 2015 March 14. Received 2015 March 11; in original form 2015 February 18}

\begin{document}
\maketitle

\newcommand{\Renaud}{Renaud et al.}


\begin{abstract}
The growth of galaxies through adiabatic accretion of dark matter is one of the main drivers of galaxy evolution. By isolating it from other processes like mergers, we analyse how it affects the evolution of star clusters. Our study comprises a fast and approximate exploration of the orbital and intrinsic cluster parameter space, and more detailed monitoring of their evolution, through $N$-body simulations for a handful of cases. We find that the properties of present-day star clusters and their tidal tails differ very little, whether the clusters are embedded in a growing galactic halo for 12 Gyr, or in a static one.
\end{abstract}
\begin{keywords}galaxies: star clusters --- methods: numerical\end{keywords}


\section{Introduction}

Since the time of their formation, galaxies have undergone a variety of transformations, from major mergers to slow accretion of dark matter and intergalactic gas \citep{Press1974, White1978, LeFevre2000, Dekel2009}. As old objects evolving in these galaxies, globular clusters and dwarf galaxies can potentially probe such changes and record some of their signatures \citep{Krauss2003, Kravtsov2005}. It is thus important to understand how the evolution of galaxies has impacted present-day stellar populations. Ideally, theoretical studies on this topic should account for the coupling between the internal dynamics of small stellar systems (clusters and dwarfs), and the effect of their galactic environment in cosmological context. However, because of the wide range of scales and of physical processes involved, considering both aspects simultaneously can be challenging. 

One possible approach consists in focussing on low-density objects like dwarf galaxies, where star-star interactions can be neglected. By adopting a collisionless treatment of gravitation in numerical simulations, one can efficiently explore a wide parameter space. Such methodology is largely used in the studies of satellite galaxies and tidal streams \citep[e.g.][]{Penarrubia2005, Bovy2014, Bonaca2014, Kuepper2015}.

But for denser systems like star clusters, two-body relaxation plays a paramount role in setting the flux of stars escaping the cluster, and thus in the evolution of the cluster \citep{Fukushige2000}. Therefore, gravitation must be treated in a collisional fashion, which involves an important numerical cost, and forbids to do so over galactic and cosmological scales \citep[see also][]{Dehnen2014}. Therefore, the large scale influence is not (yet) considered self-consistently in star cluster $N$-body simulations. Up to now, several paths have been followed to implement the tidal effects from the galaxy on a cluster:
\begin{enumerate}
\item the cluster remains within a single galaxy for its entire lifetime \citep[e.g.][]{Baumgardt2003, Kuepper2010a, Hurley2012, Madrid2012, Vesperini2014, Webb2014}. The galactic potential is usually static, and often axisymmetric.
\item the cluster is accreted from a satellite galaxy onto a massive galaxy. The accretion event is often modelled by replacing one galactic potential with another, both static \citep{Miholics2014, Bianchini2015}.
\item the galactic potential is evolved using a separate galaxy simulation to allow for time-evolving, non-analytical potentials and include effects like galaxy mergers \citep{Renaud2013a} or even complex tidal histories in cosmological context \citep{Rieder2013}.
\end{enumerate}

Although the last approach provides a relatively high level of realism, it does not allow us to distinguish the relative role of the numerous physical processes involved in the evolution of the clusters. In this paper, we focus on a specific aspect of the galactic tides, namely the secular, adiabatic cosmological growth of galaxies, and neglect other effects, at the expense of realism. The full story of the co-evolution of star clusters and their hosts would be told by considering a combination of this particular effect, and the other mechanisms driving galaxy evolution, like minor and major mergers, and the accretion of gas. In that respect, cosmological simulations and merger trees would serve as a base to establish the relative weight and frequency of these events, and to allow us to build the true tidal history of clusters, in future studies.

Here, we follow the evolution of clusters in a time-dependent galactic potential, and compare to that in static potentials, to quantify the role of secular galaxy evolution on the present-day properties of star clusters. Our method is two-fold: (\emph{i}) an exploration of the parameter space performed with very fast codes, but at the price of some simplifications, and (\emph{ii}) several much slower but more accurate $N$-body simulations using the relevant sets of parameters identified in the first step.

\section{Methodology}

\subsection{Time-evolving potential}
\label{sec:potential}

For simplicity and to limit degeneracy in the results, we only consider the halo component of the galaxy and opt for a self-similar growth. The galaxy does not experience any merger event but slowly and smoothly grows (in mass and radius) with time. Furthermore, the halo remains spherically symmetric and we neglect dynamical friction. We choose to make such simplifications in order to focus on the role of a single physical process (namely the adiabatic cosmological growth).

We use the analytical description of a growing \citet*{Navarro1997} halo proposed by \citet{Buist2014}, who fitted the evolution of halo parameters using the Aquarius simulation \citep{Springel2008}. Namely, the mass-scale and scale-length of the halo evolve with redshift $z$ as
\begin{equation}
\left\{\begin{array}{lcl}
M_s(z) & = & M_{s,0} \exp{\left( -0.2 z \right)}\\
R_s(z) & = & R_{s,0} \exp{\left( -0.1 z \right)}.
\end{array} \right.  
\label{eqn:scale}
\end{equation}
We have adopted the values of 0.1 for the mass growth parameter ($a_g$) and of 2.0 for the $\gamma$ parameter in \citet{Buist2014}, their equations 22 and 23. This provides an evolution comparable to that of the Milky Way-like halo labelled Aq-E-4 in \citet[their Fig. 4]{Buist2014}. The effect of changing these parameters is discussed in \sect{discussion}. The galactic potential, as a function of radius $r$ and redshift $z$ is then
\begin{equation}
\phi_\textrm{G}(r,z) = -\frac{GM_s(z)}{r}\ln{\left( 1 + \frac{r}{R_s(z)} \right)}
\label{eqn:potential}
\end{equation}
(where $G$ is the universal constant of gravitation), and corresponds to the density profile:
\begin{equation}
\rho_\textrm{G}(r,z) = \frac{M_s(z)}{4\pi r \left[r + R_s(z)\right]^2}.
\end{equation}

To evaluate the paramaters of the potential during the simulation of the cluster, we convert the time $t$ since the beginning of the simulation (i.e. the age of cluster) into redshift using 
\begin{equation}
1+z = \left(\frac{1-\Omega_m}{\Omega_m}\right)^{1/3} \left\{ \sinh{\left[ \left(t + t_0\right) \frac{3 H_0 \sqrt{1-\Omega_m}}{2} \right]} \right\}^{-2/3},
\label{eqn:redshift}
\end{equation}
(which follows from equation 13.20 of \citealt{Peebles1993}). $t_0$ represents the age of the Universe when the simulation is started. We adopt the values of the cosmological constants $\Omega_m = 0.31$ and $H_0 = 68 \kms\Mpc^{-1}$ \citep{Planck2014} such that the Hubble time equals $13.7 \Gyr$.

\begin{figure}
\includegraphics{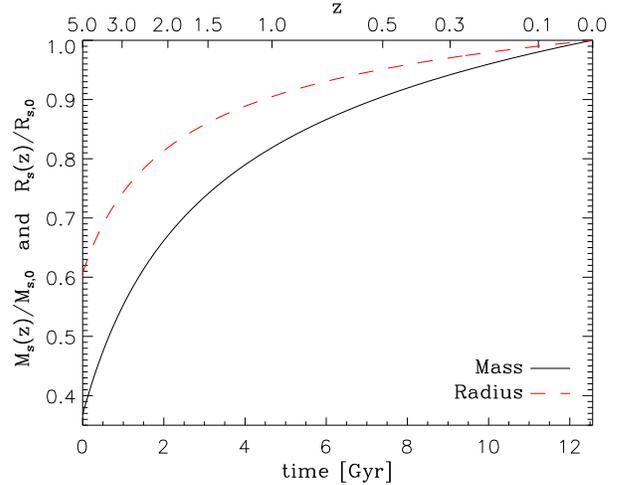}
\caption{Evolution of the scale parameters of the galaxy, normalised to their value at $z=0$ (i.e. for $t\approx 12.6 \Gyr$). $t=0$ corresponds to the start of the simulations (i.e. $z=5$).}
\label{fig:galaxy_scales}
\end{figure}

We start our study at redshift 5, which leads to $t_0 \approx 1169 \Myr$, and we consider the evolution of clusters over $12.56 \Gyr$ (corresponding to the time difference between $z=5$ and $z=0$. We choose the present day ($z=0$) values of the mass-scale and scale-length to be $M_{s,0} = 1.5 \times 10^{11} \Msun$ and $R_{s,0} = 16 \kpc$. \fig{galaxy_scales} shows the evolution of the mass-scale ($M_s$) and scale-length ($R_s$), both normalised to their values at $z=0$. 

\subsection{Galaxies}
\label{sec:galaxies}

Our study of the effects of time-varying potentials on the evolution of star cluster comprises three types of tidal histories:
\begin{itemize}
\item S5 (static, $z=5$): the cluster evolves in a static version of the potential of Equation~(\ref{eqn:potential}), with properties corresponding to $z=5$. Equation~(\ref{eqn:scale}) provides the corresponding galactic mass scale and scale radius: $\approx 5.5 \times 10^{10} \Msun$ and $\approx 9.7 \kpc$.
\item TD (time dependent): the initial setup is strictly the same as for S5, but the potential evolves with redshift from $z=5$ to $z=0$, as described in the previous Section.
\item S0 (static, $z=0$): the cluster reaches the exact same \emph{final} orbital position and velocity than in TD, but has evolved in a static potential with properties for $z=0$. In practice, we compute the final position and velocity of the cluster from TD, when $z=0$ has been reached. We then ``freeze'' the potential, and perform a backward integration of the cluster orbit. This gives us the initial position and velocity of the cluster for the S0 run.
\end{itemize}

Therefore, by setting one initial position and velocity for a cluster in the S5 galaxy, we uniquely define three orbits and the associated three tidal histories: S5, TD and S0. 

For simplicity, we always initially set the cluster at apocenter, with a purely tangential velocity, in our S5 and TD cases. Orbit integration is performed using the \nbtt method \citep{Renaud2015b}, either to integrate the motion of the cluster (Section~\ref{sec:paramspace}), or in the full $N$-body context where the cluster is described star-by-star (Section~\ref{sec:nbody}).

\subsection{Star cluster fiducial initial conditions}
\label{sec:ic}

We considered several intrinsic and orbital initial conditions for the clusters. Unless otherwise mentioned, our clusters are modeled with 32768 stars distributed on a Plummer sphere with a virial radius of $3 \pc$ (i.e. a half-mass radius of $\approx 2.3 \pc$). The masses of the stars follow a \citet{Kroupa2001b} initial mass function, from 0.1 to $1 \Msun$, leading to a total initial mass of $\approx 1.03 \times 10^4 \Msun$. We do not account for stellar evolution. We explore other cases by varying these parameters in the next Sections.

\section{Parameter space exploration}
\label{sec:paramspace}

We aim to determine if, under which conditions and to what extend, the evolution of a cluster in a cosmologically growing potential (TD) differs from that of the same cluster in a static $z=0$ version of this potential (S0). This difference relies on (\emph{i}) the strength of the tidal field and (\emph{ii}) how sensitive the cluster is to tides.

\subsection{Quantifying the tidal field}

We first seek an estimate of the galactic disruptive effect on star clusters. Such effect comprises contributions of gravitational (inertial) and orbital origin (non-inertial). For non-circular orbits, non-inertial effects, like the centrifugal force, do not yield an analytical expression. However, we can estimate their contributions by considering that the orbit is instantaneously circular, i.e. by neglecting the Coriolis and Euler effects, and by computing the centrifugal effect using the instantaneous orbital angular frequency. In this idealised framework, \citet{Renaud2011} provides the expression of the effective tidal tensor, which encompasses both the gravitational and centrifugal effects. Using the expression of the galactic potential (Equation~\ref{eqn:potential}), the main eigenvalue of the effective\footnote{The main eigenvalue of the purely gravitational tidal tensor (i.e. neglecting the centrifugal effect) is:
\begin{equation}
\lambda(r, z) = \frac{G M_s(z)}{r^3} \left\{ 2 \ln{\left(1+\frac{r}{R_s(z)}\right)} - \frac{3 r^2 + 2 r R_s(z)}{\left[r+R_s(z)\right]^2} \right\}.\nonumber
\end{equation}} tidal tensor reads
\begin{equation}
\lambda_\textrm{e}(r, z) = \frac{G M_s(z)}{r^3} \left\{ 3 \ln{\left(1+\frac{r}{R_s(z)}\right)} - \frac{4 r^2 + 3 r R_s(z)}{\left[r+R_s(z)\right]^2} \right\}.
\label{eqn:lambda}
\end{equation}
This quantity represents the galactic effect along the galaxy-cluster axis, and can be used to estimate the tidal radius\footnote{In the textbook context of circular orbits around point-masses, using the effective eigenvalue leads to the definition of the tidal radius of \citet{King1962} or \citet{Binney2008}, while the purely gravitational eigenvalue corresponds to the definition of \citet{Spitzer1987}.} \citep[see][for details]{Renaud2011}. 

\begin{figure*}
\includegraphics{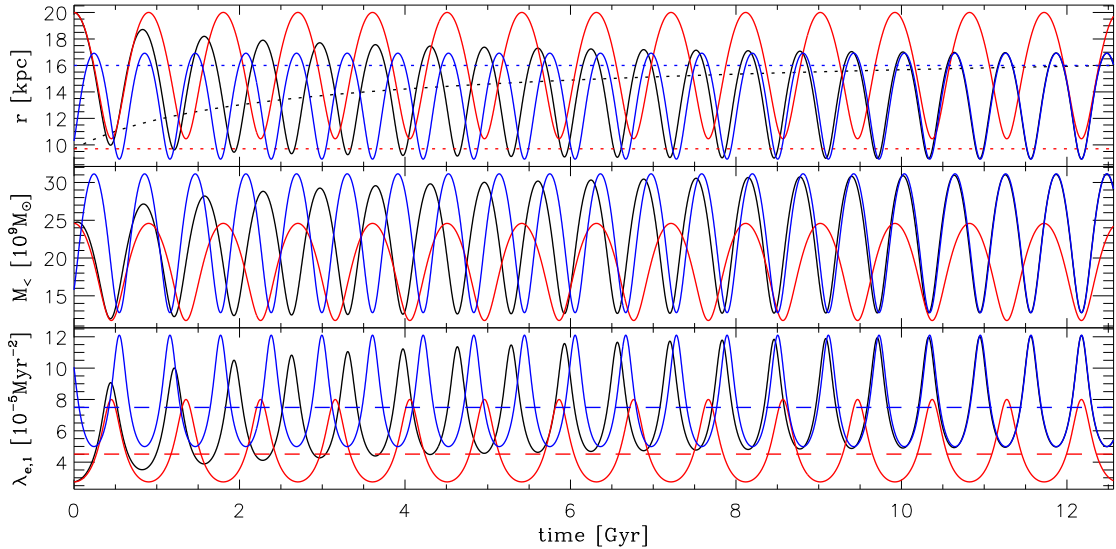}
\caption{Top: galactocentric radius for the three orbits (S5 in red, TD in black and S0 in blue). For the S5 and TD runs, the cluster is initially set $20 \kpc$ from the galactic center, with a purely tangential velocity of $50 \kms$. The initial conditions for the S0 runs are constructed as explained in \sect{galaxies}. The dotted lines indicate the corresponding scale radius ($R_s$) of the galactic potential. Middle: galactic mass enclosed within the instantaneous orbital radius. Bottom: main eigenvalue of the effective tidal tensor, under the approximation that the orbit is instantaneously circular (see text, Equation~\ref{eqn:lambda}). The dashed lines indicate the time-average of the eigenvalue along the S5 and S0 orbits.}
\label{fig:orbit}
\end{figure*}

We show an example of evolution of the effective eigenvalue in \fig{orbit}, together with the galactocentric radius and the galactic mass enclosed in this radius, for the three orbits (TD, S5 and S0). As constructed, the TD case evolves from the S5 to the S0 setup. On average, despite of a smaller orbital radius, the mass enclosed within the galactocentric radius in S0 is larger than that in S5, because the S0 galaxy yields a higher total mass. We estimate the mean tidal strength over an orbit by computing the time-average $\left<\lambda_\textrm{e}\right>$ of the main effective eigenvalue over an orbital period (dashed lines on \fig{orbit}). This quantity is constant in static potentials. The larger enclosed mass and smaller galactocentric distance result in a stronger tidal effect along the S0 orbit than in the S5 case.

\begin{figure}
\includegraphics{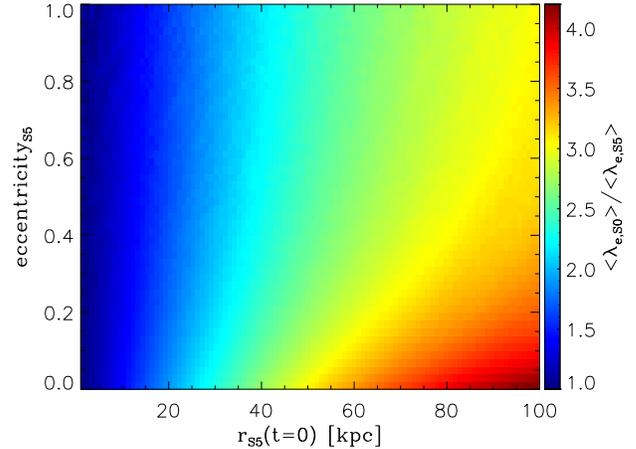}
\caption{Relative strength of the tidal field between the S0 and S5 cases, as a function of initial galactocentric radius ($r_\textrm{S5}(t=0)$) and orbital eccentricity in the S5 case (see text for details). The difference between tidal strengths is maximal for circular orbits in the outer region of the halo.}
\label{fig:eigen}
\end{figure}

The secular cosmological galactic growth can only affect the evolution of star clusters if the tidal effects in the S5 and S0 case are significantly different, or in other words, if the ratio
\begin{equation}
\frac{\left<\lambda_\textrm{e,S0}\right>}{\left<\lambda_\textrm{e,S5}\right>}
\end{equation}
is large with respect to unity. This ratio depends on the initial position and velocity of the cluster in a non-trivial way. By varying the initial galactocentric radius and orbital eccentricity of the cluster for the S5 orbit and integrating it numerically, we obtain the map of the ratio of tidal strengths shown in \fig{eigen}. (Recall from \sect{galaxies} that setting the initial position and velocity for S5 uniquely defines the initial position and velocity for S0.)

The largest differences between the tidal fields of the redshift 5 and 0 galaxies are found in the outer regions of the galactic halo, and for circular orbits. Secular galactic growth induces the largest differences in the outer regions of galactic halos and thus, the largest differences in tides are found along orbits that remain in such regions for the largest fraction of their period.

However, at large galactocentric distance, tidal forces are weak and only have a mild impact on the evolution of clusters. The differences found between the S5 and S0 cases might thus not translate into differences in the properties of the clusters.

To summarise, we have identified the orbits favouring the largest differences in tides between high and low redshift, but the resistance of clusters to tidal harassment must also be considered before concluding on the effect of secular galactic growth on star clusters.

\subsection{Cluster sensitivity to tidal harassment}

We have seen in the previous Section that the average strength of the tidal field experienced by a cluster could increase by a factor of a few because of the secular cosmological growth of the galaxy. We focus here on star cluster responses to such differences.

One of the most accurate manner to study star cluster evolution is through $N$-body simulations. These are however very numerically costly and such approach forbids an exploration of a wide parameter space. At first order however, cluster evolution relies on an handful of coupled differential equations \citep[see e.g.][]{Ambartsumian1938, Chandrasekhar1942, King1958, Henon1961, Heggie1975, Hut1992, Lee1987, Gieles2011b}. The code \emacss \citep{Alexander2012,Gieles2014,Alexander2014} solves these equations numerically and provides an easy and very fast way to evaluate the properties of the cluster along its evolution.

We first setup our fiducial cluster (\sect{ic}) in \emacss, using the tidal field strengths computed in the previous Section. To model the tidal effect, we determine the mass of the point-mass galaxy that would lead to the same average effective eigenvalue $\left<\lambda_\textrm{e}\right>$ than our NFW halos, at a given orbital radius. Doing this allows us to set the same tidal radii in \emacss than the time-average ones measured in the previous Section. However, we neglect the differences between the shape of the potentials \citep[see][for a discussion on that matter]{Tanikawa2010}. This assumption will be validated later.

\begin{figure}
\includegraphics{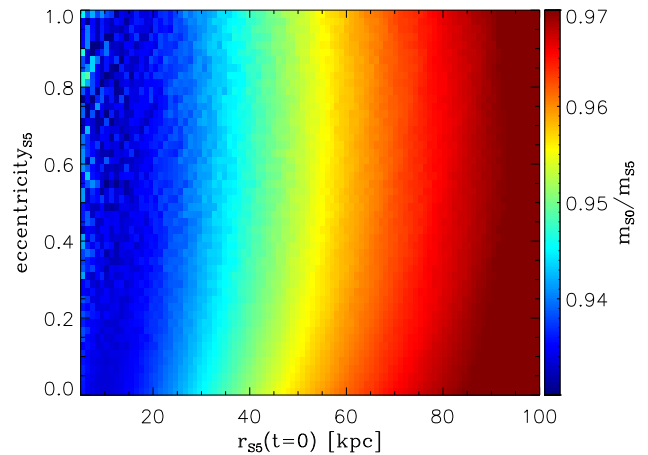}
\caption{Ratio of the final cluster mass (computed with \emacss, see text) when evolved in the S0 and S5 potentials, as a function of initial galactocentric radius and orbital eccentricity in the S5 case. The cluster initial parameters are those of the fiducial case. Note that \emacss is not designed to treat orbits with a high Roche filling factor, and thus the cases within the central $5 \kpc$ are not considered here.}
\label{fig:finalmass}
\end{figure}

\fig{finalmass} shows the ratio between the final cluster mass when embedded in the S0 and S5 tides. Despite the differences in tidal strengths found in \fig{eigen}, the differences in the final mass of the cluster remain below 10 percent in all cases. In other words, the final mass of our fiducial cluster is almost the same in S5 and S0 for all initial orbital radii and eccentricities considered.

\begin{figure}
\includegraphics{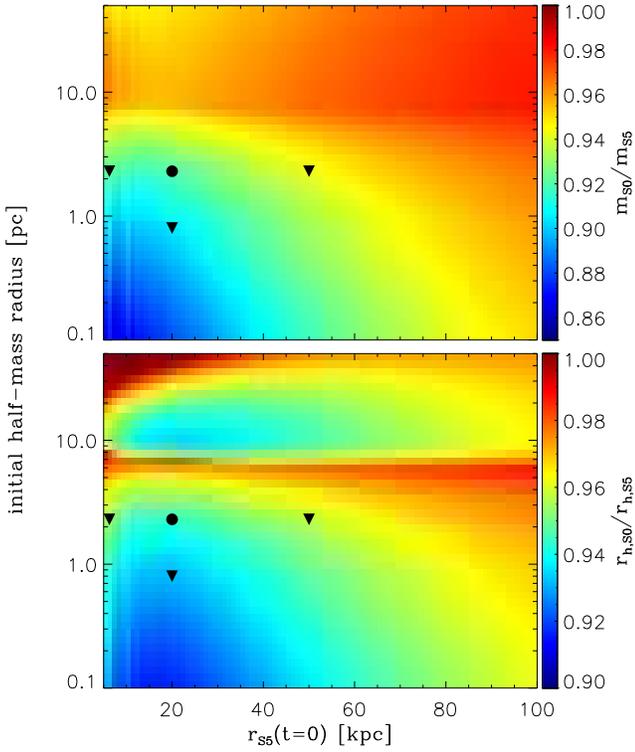}
\caption{Top: ratio of the final cluster mass when evolved in the S0 and S5 potentials, as a function of initial orbital radius (with an eccentricity of 0.3) and initial half-mass radius. The black symbols correspond to the $N$-body simulations described in \sect{nbody}. Bottom: same but for the final half-mass radius. The horizontal ridge visible in this plot is an effect of clusters reaching the core-collapse phase shortly before the end of the simulation ($\approx 12.6 \Gyr$). Clusters with a long relaxation time (large half-mass radius, above the ridge) are still in the unbalanced pre-core-collapse phase, with a decreasing half-mass radius, when the \emacss simulations are stopped \citep[see][for details]{Gieles2014}.}
\label{fig:filling}
\end{figure}

To extend this conclusion, we apply the same method to clusters with different sensitivity to tides by varying the initial half-mass radius and thus, indirectly, the Roche-filling factor (the ratio between the cluster half-mass radius and the tidal radius). We plot in \fig{filling} the ratios of the final masses (as in \fig{finalmass}) and final half-mass radii. We arbitrarily choose the tidal field strength of orbits with an eccentric of 0.3, but reach the same conclusion of other values, as already suggested by the independence to eccentricity showed in \fig{finalmass}. In all cases explored, the properties of the cluster (mass, size) along the S5 orbit lie within less than a few percent from those along the S0 orbit.

Because the TD case represents a slow and smooth transition between S5 and S0, the differences between the latter can be seen as an upper limit of the expected differences between the time-dependent (TD) and static (S0) cases. Since this upper limit is very small, we expect clusters with a time-dependent tidal history to share very similar properties as those in static potentials. In other words, cluster evolution should be fairly independent of the evolution of its host galaxy (for the secular growth considered here).

\section{$N$-body simulations}
\label{sec:nbody}

When performing the exploration of the parameter space described in the previous Section, we made the assumption that the time-average tidal strength can be used to infer the galactic influence on star clusters. While this is perfectly valid for circular orbits, it is not for eccentric cases \citep{Baumgardt2003, Webb2014}. An analytical description of tidal effects for non-circular orbit is yet to be established. Furthermore, we have neglected the second order effect of the detailed shape of the tidal field when assuming the tidal radius is a good representation of the tidal strength.

To test these assumptions and validate our conclusions, we select a few cases from our parameter space study, and perform full $N$-body simulations. We use the method and implementation \nbtt presented in \citet{Renaud2015b}, based on \nbody \citep{Aarseth2003, Nitadori2012}. The method relies on a description of the galactic potential (through a numerical routine) as a function of position and time. Using this definition, the code integrates the motion of the cluster around the galaxy, and adds tidal acceleration to the internal acceleration for all stars. We have defined the galactic potentials and their time-dependence exactly as describe in \sect{potential} and performed several sets of S5, TD and S0 runs.

\subsection{Cluster structure}

\begin{figure}
\includegraphics{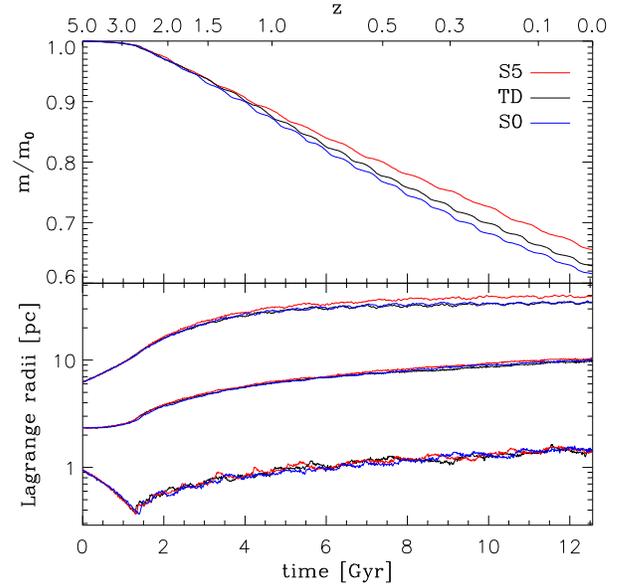}
\caption{Evolution of the mass (top) and the Lagrange radii (10, 50 and 90 percent of the mass) of our fiducial cluster along the orbits showed in \fig{orbit}, computed with \nbtt. Only stars with a negative energy (kinetic + potential from other cluster members) ar considered here, as in \citep{Renaud2015b}.}
\label{fig:fiducial}
\end{figure}

\fig{fiducial} shows the evolution of the mass and some Lagrange radii of our fiducial cluster (\sect{ic}) set at an initial position of $20 \kpc$ with an orbital eccentricity of 0.3 (as in \fig{orbit}). As expected from the parameter space exploration presented above (black dot in \fig{filling}), the relative difference in final mass (respectively half-mass radius) between the S5 and S0 runs is very small ($\approx 6.1$ percent, respectively $\sim 4$ percent). The excellent agreement between the \nbtt results and \emacss validates our assumptions, at least in this region of the parameter space. We have also tested this agreement by considering other eccentricities (0, 0.5 and 0.7, not shown here) and reached the same conclusions. As foreseen in the previous Section, the difference between the S0 and TD runs is even smaller ($\approx 2.6$ percent for the mass and $\sim 3$ percent for the half-mass radius) than that between the S5 and S0 cases.

We also run other \nbtt models, corresponding to the black triangles on \fig{filling} and reach the same conclusions, both on the validity of our method, and on the physical results obtained.

\subsection{Tidal tails}

Tidal ejection of stars from clusters leads to the formation of tidal tails. Although we have found that the mass loss rate of clusters appear to be independent of tidal history (in the context of secular growth), tidal tails could respond differently to a time-dependent and a static galactic potential.

\begin{figure}
\includegraphics{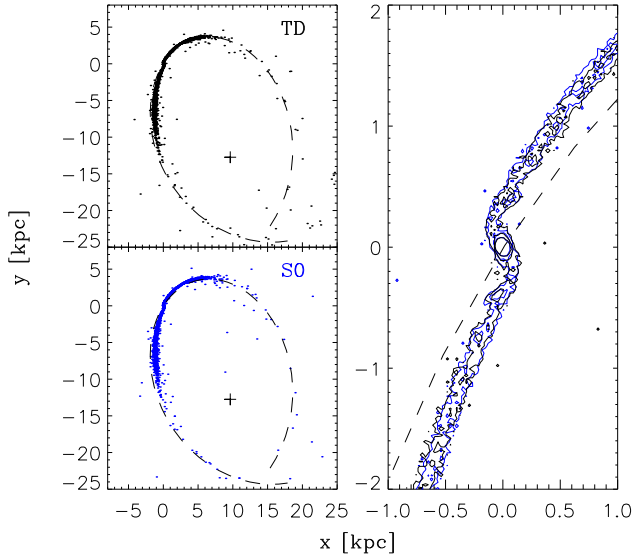}
\caption{Left: Position of the stars at redshift 0 for the TD (black) and S0 (blue) runs. The plus sign indicates the position of the center of the galactic halo. Right: Isodensity contours of the clusters and the tidal tails. The dashed line represents the orbit of the cluster. The two models overlap almost perfectly.}
\label{fig:tails}
\end{figure}

\begin{figure}
\includegraphics{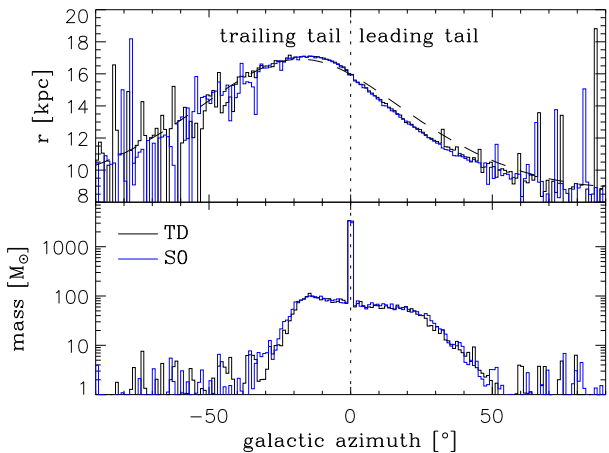}
\caption{Distribution of mean galactocentric distance of stars (top) and of the mass (bottom), as a function of azimuth in the galactic reference frame, for clusters having evolved $\approx 12.6 \Gyr$ in the TD and S0 potentials. (The cluster are centered on the azimuth $0^{\circ}$.) The dashed line represents the orbit of the cluster.}
\label{fig:tails2}
\end{figure}

\fig{tails} shows the position of all the stars (cluster and tails), and \fig{tails2} displays the distribution of mean galactocentric distance and the mass, in galatic azimuth bins (of width $1^{\circ}$), for the clusters having experienced the TD and S0 tidal histories. Differences between the models appear at rather large distance from the cluster center. The differences in mass are of the order of a few solar masses, i.e. concern only an handful of stars.

Despite different tidal histories over the $12.6 \Gyr$ of evolution we considered, the final position, density, length and distribution of substructures in the tidal tails between the TD and S0 models are almost undistinguishable. In other words, clusters and their tidal debris do not retain signatures of the tidal history they experienced.

\section{Discussion and conclusions}
\label{sec:discussion}

We study the evolution of star clusters embedded in cosmologically growing galaxies, and focus on the secular, adiabatic growth of dark matter halos, neglecting impulsive and transient events like galaxy mergers. Our main findings are:
\begin{itemize}
\item Although the typical tidal fields associated with high and low redshift halos can vary by a factor of several, these variations do not translate into major differences in the properties of star clusters embedded in those fields.
\item Because of these similarities, the details on the secular evolution of the halo (growth rate, growth epoch) have a negligible effect on the present-day properties of the clusters.
\item Present-day star clusters that co-evolved with their host galaxy yield the same properties as if they had evolved in a static halo.
\end{itemize}

Clusters orbiting in the innermost regions of galaxies are affected by the tidal effect from the baryonic components that we have neglected. Among other effects, the details on the formation of thin and thick discs could modify the role of disc shocking in the evolution of clusters \citep{Spitzer1958} and alter our conclusions. We still understand too little on galaxy and structure formation to reach conclusion on this matter.

Furthermore, we only considered spherically symmetric halos, i.e. neglecting (time-dependent) anisotropy and substructures. Although we expect these aspects to rarely be of significant importance for star clusters, \citet{Bonaca2014} showed that could alter the morphology of tidal debris.

Our conclusions depend on the growth history of the galaxy, mainly when and how fast the bulk of the growth takes place. Cosmological simulations indicate that most of the adiabatic growth happens at high redshift (recall \fig{galaxy_scales}, see \citealt{Buist2014}), which has a very mild effect on star clusters, as we have shown. A galaxy can also experience a significant growth in the form of a violent event. If so, the growth enters the impulsive regime and can be seen a galaxy merger. The response of the clusters to such event is non-negligible but complex, as showed in \citet{Renaud2013a}. The real evolution of a galaxy and of its clusters lies in between these two extremes. Our study demonstrates that the main impacts of galactic tides on star clusters in the cosmological context are the impulsive events, and not the adiabatic growth.

In the context of the Milky-Way, the absence of evidence for recent major mergers (in the last $\sim 6-9 \Gyr$, \citealt{Deason2013}, and/or since the formation of the disc \citealt{Ruchti2014}) makes our results directly applicable to clusters formed \emph{in-situ}. Studies of such clusters can thus safely make the assumption that the galactic potential has been static over several Gyr, without biasing the conclusions on the mass- and size-functions of the clusters. Studies that consider globular cluster population evolution in a static Milky Way potential showed that it is not possible, within a Hubble time, to evolve an initial cluster mass function with a power-law shape and an index of -2 (as found for young massive cluster today, \citealt{Portegies2010}), into a mass function that is peaked at a universal mass of $\sim 2 \times 10^5 \Msun$ with secular dynamical evolution \citep{Baumgardt1998, Vesperini2001, Gieles2008b}. Our results support this conclusion. Note however that this is not valid when considering the non-negligible fraction of clusters of external origin, which have been accreted from dwarf satellite galaxies onto the Milky Way \citep{Marin-Franch2009, Leaman2013}.

The diversity of scenarios of galaxy evolution seen in cosmological simulations implies that present-day star clusters have experienced a wide variety of tidal histories. By decomposing these scenarios into individual processes and events like major mergers \citep{Renaud2013a}, accretion of dwarf satellites \citep{Bianchini2015}, secular growth (this work) and others, and by understanding their relative roles on the evolution of clusters, we will soon able to seek and identify specific imprints of galaxy evolution on the observed properties of star clusters.

\section*{Acknowledgements}
We thank Hans Buist for his help in choosing the halo parameters, and the reviewer for a prompt and constructive repport. We acknowledge support from the European Research Council through grant ERC-StG-335936 (CLUSTERS). MG acknowledges financial support from the Royal Society in the form of a University Research Fellowship and an equipment grant that was used to purchase the GPU machines that were used for the $N$-body computations. 
\bibliographystyle{mn2e}

\end{document}